# DMRF-UNet: A Two-Stage Deep Learning Scheme for GPR Data Inversion under Heterogeneous Soil Conditions

Qiqi Dai, Yee Hui Lee, *Senior Member, IEEE*, Hai-Han Sun, Genevieve Ow,
Mohamed Lokman Mohd Yusof, and Abdulkadir C. Yucel, *Senior Member, IEEE*

*Abstract*—Traditional ground-penetrating radar (GPR) data inversion leverages iterative algorithms which suffer from high computation costs and low accuracy when applied to complex subsurface scenarios. Existing deep learning-based methods focus on the ideal homogeneous subsurface environments and ignore the interference due to clutters and noise in real-world heterogeneous environments. To address these issues, a two-stage deep neural network (DNN), called DMRF-UNet, is proposed to reconstruct the permittivity distributions of subsurface objects from GPR B-scans under heterogeneous soil conditions. In the first stage, a U-shape DNN with multi-receptive-field convolutions (MRF-UNet1) is built to remove the clutters due to inhomogeneity of the heterogeneous soil. Then the denoised B-scan from the MRF-UNet1 is combined with the noisy B-scan to be inputted to the DNN in the second stage (MRF-UNet2). The MRF-UNet2 learns the inverse mapping relationship and reconstructs the permittivity distribution of subsurface objects. To avoid information loss, an end-to-end training method combining the loss functions of two stages is introduced. A wide range of subsurface heterogeneous scenarios and B-scans are generated to evaluate the inversion performance. The test results in the numerical experiment and the real measurement show that the proposed network reconstructs the permittivities, shapes, sizes, and locations of subsurface objects with high accuracy. The comparison with existing methods demonstrates the superiority of the proposed methodology for the inversion under heterogeneous soil conditions.

## I. Introduction

GROUND-penetrating radar (GPR) has been widely used to detect subsurface objects and image the subsurface structures in geophysical and civil engineering applications. In GPR inverse problems, the properties of the subsurface scenario can be recovered from the electromagnetic (EM) information in the recorded GPR data. Reconstructing the permittivity distributions of subsurface objects, including constitutive parameters, locations, sizes, and shapes, from the B-scans, has practical significance in the non-destructive health monitoring and subsurface utility mapping [1], [2].

There are several classical methods for reconstructing subsurface structural images from the acquired GPR B-scans, including migration algorithms, tomographic approaches, and full-waveform inversion (FWI) techniques. Migration algorithms, such as the reverse-time migration [3] and the Kirchhoff migration [4], have been developed for transforming the unfocused space-time GPR image to a focused one, showing the object's location and size. However, such algorithms do not provide the constitutive parameters of the object, of great importance for identifying subsurface objects and monitoring their conditions. Microwave tomography approaches [5], [6] are investigated to solve inverse scattering problems and reconstruct the subsurface dielectric profiles from the acquired GPR data. However, the requirements of the measurement configuration are often high, and the iterative inversion process is computationally expensive. FWI algorithms [2], [7], [8] are proposed to reconstruct subsurface permittivity maps by mapping the subsurface structure information from GPR data via a non-linear least-squared optimization process. However, the forward modeling needs to be performed in every iteration, while the optimization usually requires a large number of iterations. This results in high computational cost. Besides, the reconstruction accuracy is often low when these algorithms are applied to complex subsurface scenarios. In short, the limitations of the conventional methods are listed as follows: (i) They are not capable of restoring the objects' constitutive parameters, or (ii) they require high computational cost when iteratively constructing the permittivity maps, and (iii) they provide inaccurate results when applied to permittivity map reconstruction of complex subsurface scenarios.

To improve the inversion efficiency and accuracy, deep learning-based methods have been investigated for the GPR applications, such as hyperbolic signature recognition, image classification, object detection, and restoration of the objects' properties [9]-[12]. To restore the subsurface permittivity distributions, deep neural networks (DNNs) have been utilized to find the relationships between GPR B-scans and permittivity maps [13]-[16]. In [13], three DNNs for image-to-image transformation, namely the encoder-decoder [17], U-Net [18], generative adversarial network (GAN) [19], were employed to reconstruct the subsurface permittivity maps of sewer crowns.

This work was supported by National Parks Board, Singapore (*Corresponding authors: Yee Hui Lee; Abdulkadir C. Yucel*).

Q. Dai, Y. H. Lee, H. -H. Sun and A. C. Yucel are with the School of Electrical and Electronic Engineering, Nanyang Technological University, Singapore 639798 (e-mails: qiqi.dai@ntu.edu.sg, eyhlee@ntu.edu.sg, haihan.sun@ntu.edu.sg, acyucel@ntu.edu.sg).

G. Ow and M. L. M. Yusof are with the National Parks Board, Singapore 259569 (e-mails: genevieve_ow@nparks.gov.sg, mohamed_lokman_mohd_yusof@nparks.gov.sg).

The testing results showed that the encoder-decoder and U-Net outperformed GAN. This study proved the possibility of transforming the non-intuitive B-scans into easy-to-interpret subsurface images. In [14], the U-Net with instance normalization was adopted to map the permittivity distribution from recorded GPR data in real-time. In [15], a trace-to-trace network GPRInvNet was proposed to deal with complex GPR B-scans of tunnel linings. To address attenuation-induced issues and extract sufficient global information, an improved network PINet [16] was proposed to reconstruct the permittivity map of geo-structures from B-scans. The comparative results on both numerical simulation and real measurement data in [15] and [16] demonstrated the superior performance of the deep learning-based methods compared to conventional algorithms. The PINet achieves state-of-the-art performance in accurately reconstructing permittivity distributions. However, all these deep learning-based studies only considered the inversion for homogeneous environments and the obtained B-scans contain little environmental clutters and noise. Only the study in [15] randomly cropped the noise patches obtained from the real heterogeneous background and added them to synthetic B-scans for network training. Yet, the reconstruction accuracy is greatly reduced, and the network can only reconstruct rough profiles of the subsurface objects. This is because the noise and clutters from the heterogeneous background interfere with the signatures corresponding to reflections from the objects and make the identification of reflection patterns in B-scans difficult, rendering the GPR inversion task more challenging in complex heterogeneous background conditions. Especially, when the object's permittivity is close to the surrounding soil or several objects are in close proximity to each other, the hyperbolic signatures of the objects will be much harder to recognize and even masked by the environmental clutters.

In this work, a two-stage DNN, named DMRF-UNet, is proposed to address the issues in the permittivity distribution reconstruction of subsurface objects under heterogeneous soil conditions. The DMRF-UNet consists of double U-shape convolutional neural networks (CNNs) with multiple receptive fields (MRF-UNet) to describe the representation of the GPR inverse problem. The first MRF-UNet (MRF-UNet1) extracts the hyperbolic signatures reflected from subsurface objects and removes the clutters and noise due to the heterogeneous soil in the input noisy B-scan. The denoised B-scan and the noisy B-scan are concatenated as input to the second MRF-UNet (MRF-UNet2). The MRF-UNet2 learns the mapping relationship between the signatures in the two-channel B-scans and the permittivity distribution of subsurface objects. In two stages, convolutions with multiple receptive fields are utilized to extract multi-scale hyperbolic signatures corresponding to the reflections from subsurface objects in the B-scans. The end-to-end training is conducted based on the combined loss from two stages to avoid information loss. A dataset including the noisy B-scans obtained under heterogeneous soil conditions, the denoised B-scans, and the permittivity distributions of subsurface objects are generated to train and test the DMRF-UNet. (Note: The noisy B-scans are directly obtained through GPR scanning under heterogeneous soil conditions rather than the addition as in [15].) The comparative results on numerical simulation and real measurement data demonstrate the superior performance of the proposed method over existing studies.

The main contributions of this work are as follows: 1) This is the first time that a deep learning-based framework is proposed for reconstructing the permittivity distributions of subsurface objects under heterogeneous soil conditions. 2) A unique two-stage DNN which outperforms existing works for the reconstruction under heterogeneous soil conditions is carefully designed. (i) A novel prior denoising network is introduced to extract the objects' signatures and eliminate the noise interference under heterogeneous soil conditions. (ii) The DNN employs multiple receptive fields to extract multi-scale hyperbolic signatures corresponding to the reflections from multiple objects with various properties in the B-scan. Cascaded small-size convolutional filters are specifically used instead of the large-size filters to reduce the computation cost. (iii) The two-stage DNN is trained end-to-end with a combined loss function to balance the training of the two stages and avoid information loss. The proposed network that combines all these components achieves the best performance for reconstructing permittivity distributions under heterogeneous soil conditions.

## II. METHODOLOGY

Let $X$ and $Y$ represent the subsurface scenario and the resulting B-scan obtained after the GPR survey performed on the surface, as shown in Fig. 1. The forward relationship between the input subsurface scenario and output B-scan is represented by $H(\cdot)$ operator and the GPR forward problem can be described as $Y = H(X)$. In the inverse problem, the aim is to find the inverse transformation $H^{-1}(\cdot)$ from $Y$ to $X$ ($X = H^{-1}(Y)$). The inverse transformation is highly non-linear and the inverse problem is ill-posed. Supervised deep learning provides a data-driven technique to learn $H^{-1}(\cdot)$ to translate the B-scan into the permittivity distribution of subsurface objects. A large set of data is required to train the DNN by minimizing the loss between the predicted and actual permittivity distributions. After the network is well-trained, which represents the optimal $H^{-1}(\cdot)$, the network can automatically predict permittivity distributions from new B-scans.

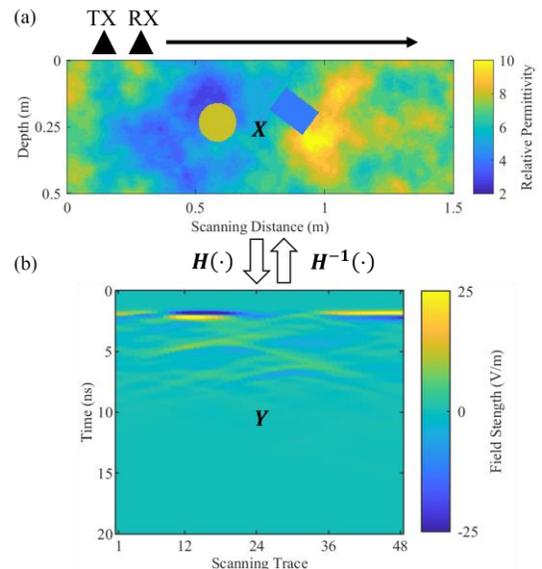

Fig. 1. (a) A typical subsurface scenario $X$ with the GPR system under the heterogeneous soil condition. (b) The obtained B-scan image $Y$.

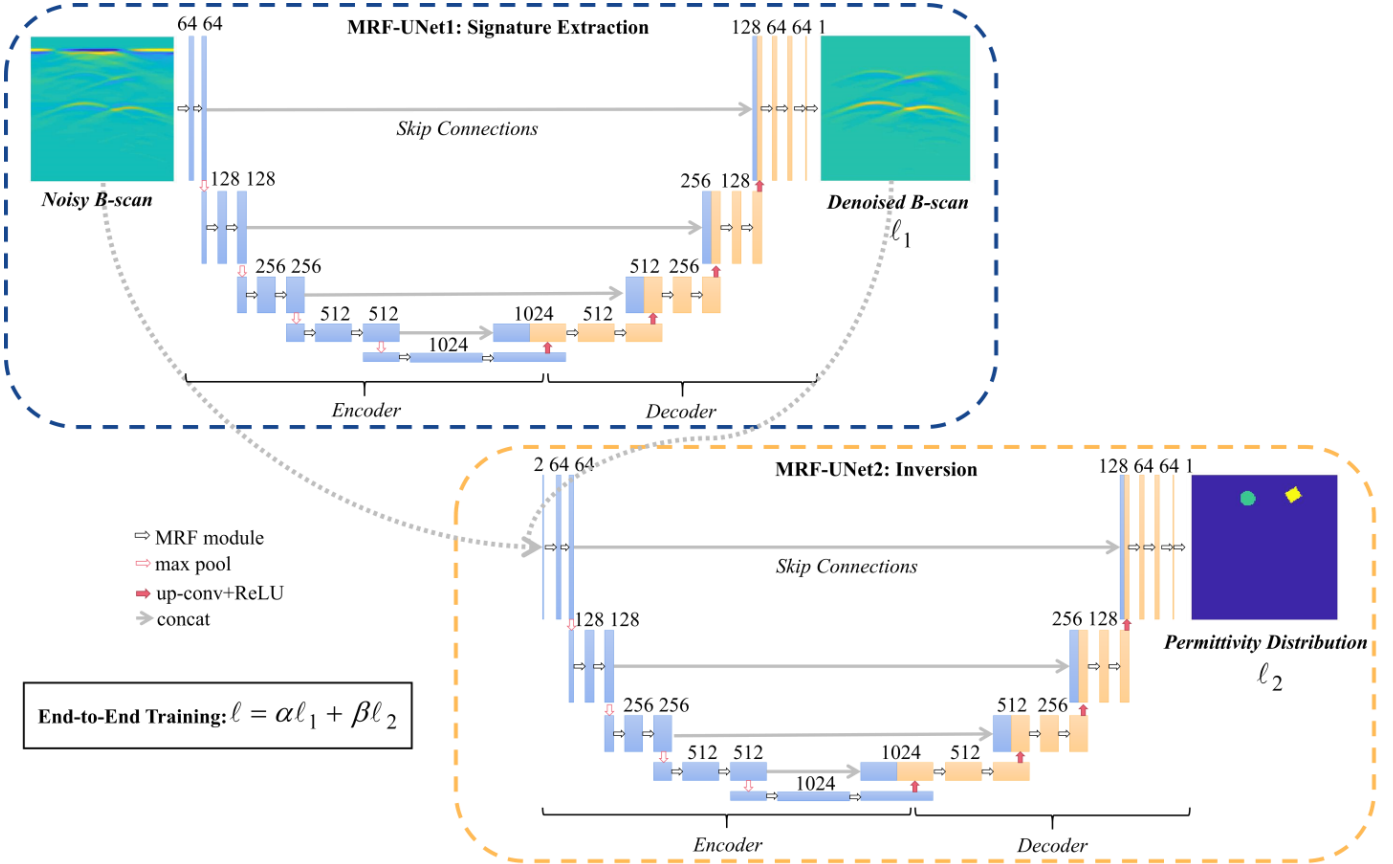

Fig. 2. The structure of the proposed DMRF-UNet. 'MRF module', 'max pool', 'concat', 'up-conv', 'conv', and 'ReLU' represent the multi-receptive-field module, max-pooling layer, concatenation, up-convolutional layer, convolutional layer, and rectified linear unit activation, respectively. The numbers above the rectangles mean the channels of feature maps. $\ell_1$ and $\ell_2$ represent the loss functions of MRF-UNet1 and MRF-UNet2. $\ell$ is the combined loss for the end-to-end training.

To reconstruct the subsurface permittivity distributions from the GPR B-scans under heterogeneous environment conditions, there exist two main challenges. First, the heterogeneity of subsurface environment introduces clutters and noise in the B-scan that interferes with signatures of reflections from the objects. Especially when the object's permittivity is close to that of the surrounding soil, the reflected field strength from the object is weak. Thereby, the object's reflection can be easily disguised by environmental clutters. Second, complex subsurface scenarios often yield complicated reflection patterns in the obtained B-scans. When several objects with different properties are close to each other, their reflected hyperbolic patterns are overlapped or obscured, which makes it hard to differentiate the objects' signatures and perform the subsequent permittivity distribution reconstruction. To solve these issues, a two-stage DNN is designed to first denoise the B-scan under heterogeneous soil conditions and then reconstruct the permittivity distributions of subsurface objects. Fig. 2 shows the whole architecture. The details are introduced as follows.

### A. MRF-UNet1: Signature Extraction

Compared with the homogeneous environment, clutters and noise are more pronounced in the B-scans under heterogeneous soil conditions, such as the input noisy B-scan shown in Fig. 2. It is imperative to first denoise the noisy image and extract the signatures corresponding to the reflections from subsurface objects to improve the quality of B-scans for further analysis. Based on the blind source separation principle in signal processing [20], it is assumed that the observed GPR data ($x$) is a mixture of the clutter ($x_c$) and the target signal ($x_t$) [21], as expressed as

$$x = x_c + x_t. \quad (1)$$

Signature extraction aims to remove the clutter component $x_c$ in the obtained noisy B-scan and to provide a denoised image that only includes the target component $x_t$.

To achieve that, a U-shape CNN with MRF modules, namely MRF-UNet1 in Fig. 2, is designed. The noisy B-scan is obtained by preprocessing the original B-scan by the mean subtraction technique [22]. MRF-UNet1 takes the noisy B-scan $x$ as input, extracts key features of the object reflection from $x$, suppresses the environmental clutters $x_c$, and outputs the denoised B-scan with only object reflections $x_t$. The network consists of an encoder and a decoder with skip connections. The encoder is made up of five repeated applications of two MRF modules and one $2 \times 2$ max-pooling layer with the stride of $2 \times 2$. The decoder includes four repeated applications of one up-convolution layer and two MRF modules. A $2 \times 2$ up-sampling layer is combined with a $2 \times 2$ convolutional layer as one up-convolution layer. The channels of the convolutional layers in the encoder and decoder are set as [64, 128, 256, 512, 1024] and [512, 256, 128, 64], respectively. To compensate for the information loss in the down-sampling process of the encoder, four skip connections concatenate the feature maps from the up-convolutional layers in the decoder with the corresponding feature maps from the encoder. At the final stage, a $1 \times 1$ convolutional layer followed by the rectified linear unit (ReLU) activation outputs the denoised B-scan, which only remains hyperbolic signatures reflected from objects.

In the training phase, the ground truth of the denoised image is generated by subtracting the simulated B-scan under heterogeneous soil conditions without any objects from the B-scan under heterogeneous soil conditions with the objects. The denoised images only have the hyperbolic signatures from the objects, as shown in the output image of the MRF-UNet1 in Fig. 2. Note that the ground truth of the prior denoised image is only required in the training phase. The well-trained network will automatically predict the denoised B-scans.

*B. MRF-UNet2: Inversion*

After extracting the object-related signatures and removing the soil-related noise, the MRF-UNet2 is introduced to accomplish the task of inversion, which translates the EM information in the B-scans into the permittivity distributions of subsurface objects. The configuration of MRF-UNet2 is shown in Fig. 2. The noisy B-scan and the denoised B-scan from MRF-UNet1 are concatenated as a two-channel input of the MRF-UNet2. This arrangement is made to allow the MRF-UNet2 to concentrate on the object-related signatures from the predicted denoised B-scan while avoiding loss of information in the B-scan due to the denoising process of MRF-UNet1. In this way, the reconstruction capability of MRF-UNet2 can be enhanced by learning more informative features from the hybrid B-scans. Using the same structure as MRF-UNet1, MRF-UNet2 is composed of an encoder and a decoder with skip connections, while the input is the hybrid two-channel B-scan and the output is the subsurface permittivity distribution. At the final stage, a $1 \times 1$ convolutional layer followed by the exponential linear unit activation outputs the permittivity distribution of subsurface objects. Other parameter settings in MRF-UNet2 are the same as those in MRF-UNet1.

*C. Multi-Receptive-Field Module*

A GPR B-scan usually includes multiple hyperbolic patterns at different spatial locations due to the randomness of the shape, location, and property of the subsurface object. To effectively extract discriminative features in different spatial scales, inspired by the Inception module in [23], a specific MRF module is designed in the convolutions. The receptive field is theoretically defined as the region in the input image that a particular CNN's feature is looking at, which means how large the pixels in the high-level feature map are affected by the input image [24]. Large receptive fields capture more global features with higher-level semantic information, while small receptive fields perceive the features that focus on the local details. The receptive field size is computed by [25]

$$r_\eta = r_{\eta-1} + (k_\eta - 1) \times \prod_{n=1}^{\eta-1} s_n, \quad (2)$$

where $r_\eta$ is the receptive field of the $\eta^{th}$ layer, $r_{\eta-1}$ is the receptive field of the $(\eta-1)^{th}$ layer, $k_\eta$ is the kernel size and $s_n$ is the stride in the convolutions. If $s_n$ is set as 1 in every convolutional layer, the receptive field is determined by the kernel size of the current layer $k_\eta$ and the receptive field of the former layer $r_{\eta-1}$. Thus, four kernel sizes, $1 \times 1$, $3 \times 3$, $5 \times 5$, and $7 \times 7$, as shown in Figs. 3(a)-(d), are designed to achieve four different receptive fields, $1 \times 1$, $3 \times 3$, $5 \times 5$, and $7 \times 7$, to extract multi-scale features in the B-scan. However, the computational cost increases with a larger kernel size. To alleviate the computational burden, the mini-network replacement scheme [26] is used, which replaces the $5 \times 5$ convolution with two cascaded $3 \times 3$ convolutional layers [Fig. 3(e)] and replaces the $7 \times 7$ convolution with three cascaded $3 \times 3$ convolutional layers [Fig. 3(f)]. The receptive field after the replacement remains unchanged based on (2). Assuming the channels of the feature map is $C$, the required numbers of trainable parameters in the $5 \times 5$ and $7 \times 7$ convolutions are $25C^2$ and $49C^2$, respectively. After applying the mini-network replacement, the numbers of parameters are reduced to $18C^2$ and $27C^2$, respectively, demonstrating the effectiveness of the replacement scheme in alleviating the computational burden. Moreover, the depth of the network increases and the network capacity and complexity are enhanced with the mini-network replacement scheme.

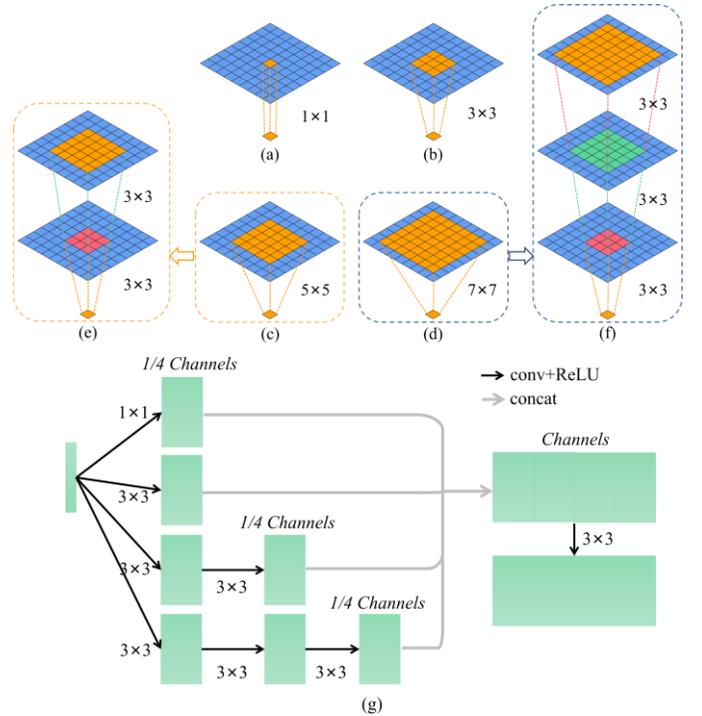

Fig. 3. Four receptive fields from four different kernel sizes: (a) $1 \times 1$, (b) $3 \times 3$, (c) $5 \times 5$, and (d) $7 \times 7$. (e) The $5 \times 5$ convolution in (c) can be replaced by two cascaded $3 \times 3$ convolutional layers. (f) The $7 \times 7$ convolution in (d) can be replaced by three cascaded $3 \times 3$ convolutional layers. (g) The structure of one MRF module. The green rectangles represent the feature maps. 'conv', 'ReLU', and 'concat' represent the convolutional layer, rectified linear unit activation, and concatenation, respectively.

The detailed structure of one MRF module is presented in Fig. 3(g). Firstly, convolutions with four sizes of receptive fields: $1 \times 1$, $3 \times 3$, $5 \times 5$ (two cascaded $3 \times 3$ convolutional layers), and $7 \times 7$ (three cascaded $3 \times 3$ convolutional layers), are conducted. Each convolutional layer is followed by the ReLU activation. The stride of each convolutional layer is $1 \times 1$. Then the obtained four feature maps are concatenated, followed by a 3×3 convolutional layer. The final number of channels becomes four times the previous. The MRF module is implemented in both the MRF-UNet1 for signature extraction and the MRF-UNet2 for inversion.

*D. End-to-End Training*

The losses of MRF-UNet1 and MRF-UNet2, $\ell_1$ and $\ell_2$, are combined in the loss function to implement end-to-end training of the proposed framework. Both the $\ell_1$ and $\ell_2$ use the mean square error (MSE) loss. The combined loss is defined as:

$$\ell = \alpha \ell_1 + \beta \ell_2 = \alpha \frac{1}{I_1 \times J_1} \sum_{i_1, j_1} \left( y_{1_{i_1,j_1}} - \hat{y}_{1_{i_1,j_1}} \right)^2 + \beta \frac{1}{I_2 \times J_2} \sum_{i_2, j_2} \left( y_{2_{i_2,j_2}} - \hat{y}_{2_{i_2,j_2}} \right)^2, \quad (3)$$

where $\ell$ is the combined loss. $\ell_1$ and $\ell_2$ are the losses of MRF-UNet1 and MRF-UNet2, respectively. When $\ell_1$ and $\ell_2$ are different in order of magnitude in the training process, the training will be biased towards the sub-network with a larger loss. To avoid this issue, the loss weights $\alpha$ and $\beta$ are used to balance the training of MRF-UNet1 and MRF-UNet2. The values of $\alpha$ and $\beta$ are determined by the quantitative difference between $\ell_1$ and $\ell_2$. $I_1$ and $J_1$ are the dimensions of the denoised B-scan, while $i_1$ and $j_1$ are the indices. $I_2$ and $J_2$ are the dimensions of the permittivity distribution, while $i_2$ and $j_2$ are the indices. $y_1$ and $\hat{y}_1$ are the ground truth and the predicted result of the denoised B-scan. $y_2$ and $\hat{y}_2$ are the ground truth and the predicted result of the permittivity distribution.

## III. NUMERICAL EXPERIMENTS

*A. Numerical Dataset Preparation*

To train and test the proposed DMRF-UNet, a large diverse set of data under heterogeneous soil conditions is generated using a GPU-based open-source software gprMax [27], [28]. Each set of data includes three images, the input noisy B-scan, the denoised B-scan, and the permittivity distribution of the subsurface environment with buried objects. To obtain the noisy B-scans under the heterogeneous soil environment with realistic dielectric and geometric properties, a Peplinski mixing model [29] implemented in gprMax is used. The properties of soil are set as sand fraction 0.5, clay fraction 0.5, bulk density 2 g/cm$^3$, and sand particle density 2.66 g/cm$^3$. The fractal dimension is set as 1.5 and the weights of the fractals along the $x$, $y$, and $z$ axes are the same. 20 different materials over a range of water volumetric fractions from 0.1% to 20%, described by 20 Debye functions, are randomly distributed in the soil. The relative permittivity and conductivity of the soil vary in [3.82, 9.99] and [0.01, 0.07], respectively. To increase the diversity of the heterogeneous soil, ten random distributions are generated to model 10 heterogeneous scenarios.

One example of the heterogeneous soil with two buried objects is shown in Fig. 1(a). The size of the 2D soil domain is 1.5×0.5 m$^2$ and the resolution is 0.0025×0.0025 m$^2$. The time window is set as 20 ns. A Gaussian waveform with an amplitude of 1 A and a center frequency of 1 GHz is excited. A Hertzian dipole (TX) is used to transmit the signals and a probe (RX), 20 cm away from the dipole, is used to receive the signals. The distance between the TX/RX and the soil surface is 10 cm. In a common-offset mode, the TX/RX are moving along one straight scanning trajectory to obtain a B-scan. The scanning step is 2.5 cm. The shape of the buried object is randomly selected from circles, semi-circles, triangles, and rectangles. The object's permittivity is randomly selected from [2, 32]. For the circle and semi-circle, the radius is randomly selected from [0.05, 0.08] m. For the triangle, the distance from the three vertices to the center is randomly selected from [0.05, 0.08] m. The $x$ and $y$ positions of the center are randomly selected from [0.25, 1.25] m and [0.25, 0.4] m, respectively. For the rectangle, the $x$ and $y$ positions of the left bottom vertex are selected from [0.5, 1] m and [0.25, 0.3] m. The width and length are chosen from [0.04, 0.06] m and [0.12, 0.16] m. The orientations of the semi-circle, triangle, and rectangle are selected from [0, 360] degrees. 8,000 one-object scenarios and 10,000 two-object scenarios are simulated to obtain the B-scans.

To obtain the denoised B-scans as the prior label of the MRF-UNet1, 10 heterogeneous environments without any objects are used as the simulation scenarios to obtain the soil-only B-scan with only background clutters. After that, the soil-only B-scans are subtracted from the noisy B-scans to create the object-only (denoised) B-scans. To generate permittivity distributions as the second label, we set the soil area as 0 and the object region as the object's permittivity. The dataset including the noisy B-scans, the denoised B-scans, and the permittivity distributions is used for training and testing purposes.

*B. Implementation Details*

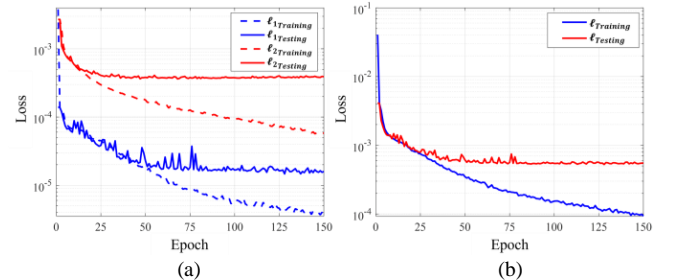

Fig. 4. The training and testing loss curves of (a) $\ell_1$, $\ell_2$, and (b) $\ell$.

After obtaining the dataset, 16200 sets of data (7200 one-object and 9000 two-object scenarios) are used for training while the rest 1800 sets of data (800 one-object and 1000 two-object scenarios) are for testing. The original B-scans are preprocessed by mean removal [22]. Then all the images are normalized to [0, 1] and resized to 128×128. The proposed DMRF-UNet is implemented on TensorFlow [30]. Note that the whole network has one input (the noisy B-scan) and two outputs (the denoised B-scan and the permittivity distribution). The end-to-end training is employed based on the loss function (3). The Adam optimizer [31] is used to perform the optimization. The learning rate is set as 0.0001. The network is trained for 150 epochs and the best model with the lowest testing loss is saved. Fig. 4(a) shows the training and testing loss curves of $\ell_1$ and $\ell_2$ of the two sub-networks, respectively. In the training loss curves, $\ell_1$ is always about 10 times larger than $\ell_2$. To balance the training of MRF-UNet1 and MRF-UNet2, $\alpha$ and $\beta$ are set as 10 and 1, respectively. Fig. 4(b) presents the combined loss curves of $\ell$ with good convergence.

Four metrics are applied to the testing data to quantitatively evaluate the inversion performance of the proposed scheme. The structural similarity (SSIM) is adopted to evaluate the similarity of the predicted result $y$ ($y_1$ or $y_2$) and the ground truth $\hat{y}$ ($\hat{y}_1$ or $\hat{y}_2$) (the larger, the better):

$$SSIM = \frac{(2\mu_y\mu_{\hat{y}} + c_1)(2\sigma_{y\hat{y}} + c_2)}{(\mu_y^2 + \mu_{\hat{y}}^2 + c_1)(\sigma_y^2 + \sigma_{\hat{y}}^2 + c_2)}, \quad (4)$$

where $\mu_y$ and $\mu_{\hat{y}}$ are the means of $y$ and $\hat{y}$. $\sigma_y$ and $\sigma_{\hat{y}}$ are the variances of $y$ and $\hat{y}$, respectively. $\sigma_{y\hat{y}}$ is the covariance of $y$ and $\hat{y}$. $c_1 = (0.01R)^2$ and $c_2 = (0.03R)^2$ are two variables where $R$ is the dynamic range of the image. The MSE, mean absolute error (MAE), and mean relative error (MRE) evaluate the errors between $y$ and $\hat{y}$ (the smaller, the better):

$$MSE = \frac{1}{I \times J} \sum_{i,j} (y_{i,j} - \hat{y}_{i,j})^2, \quad (5)$$

$$MAE = \frac{1}{I \times J} \sum_{i,j} |y_{i,j} - \hat{y}_{i,j}|, \quad (6)$$

$$MRE = \frac{1}{I \times J} \frac{\sum_{I,J} |y_{i,j} - \hat{y}_{i,j}|}{max(|y_{i,j}|)} \times 100\%. \quad (7)$$

### C. Inversion Results and Comparative Study

#### C. 1. Comparison with Other Deep Learning-based Methods

The testing performance of the proposed DMRF-UNet is compared with four deep learning-based models: (1) PINet [16], which is state-of-the-art GPR inversion work based on deep learning, (2) Enc-Dec (Note: To control the variables in the comparison, we adopt the encoder-decoder structure in U-Net [18] but remove the skip connections), (3) U-Net [18], and (4) the single MRF-UNet (SMRF-UNet) without the MRF-UNet1. For all these models, the permittivity distribution of subsurface objects is predicted from the noisy B-scan directly without the inclusion of the denoising sub-network. For the models (1)-(3), the convolutions are regular ones without MRF modules. After inversely normalizing the output denoised B-scans and permittivity distributions to [-50, 75] V/m and [0, 32], respectively, the comparison on the evaluation metrics of the testing data is shown in Table I. The evaluation metrics of the first-stage denoising and the second-stage inversion results of the DMRF-UNet are listed as "#1" and "#2", respectively. As shown, the DMRF-UNet achieves the best inversion performance on all the metrics and the SMRF-UNet is the second best compared to the PINet, Enc-Dec, and U-Net. This is because the networks (1)-(3) employ single-receptive-field convolutions, which leads to a weaker feature extraction capability when facing scenarios with multiple objects buried in the heterogenous soil. On the other hand, all the networks (1)-(4) extract features only from the noisy B-scan. The clutters and noise from the heterogenous soil make the identification of the objects' hyperbolic signatures difficult, which is also verified by the SMRF-UNet's performance degradation comparing to the DMRF-UNet's. However, the DMRF-UNet includes MRF modules to extract multi-scale signatures, the denoising sub-network to first extract the objects' hyperbolic signatures, and the inversion sub-network to perform the precise reconstruction of the permittivity distribution.

TABLE I
COMPARISON WITH DEEP LEARNING-BASED STUDIES ON EVALUATION METRICS OF THE TESTING DATA

| Network | | SSIM (↑) | MSE (↓) | MAE (↓) | MRE (%) (↓) |
|---|---|---|---|---|---|
| PINet [16] | | 0.9617 | 0.8552 | 0.0563 | 0.2987 |
| Enc-Dec [18] | | 0.9794 | 0.5222 | 0.0428 | 0.2158 |
| U-Net [18] | | 0.9803 | 0.4968 | 0.0399 | 0.2039 |
| SMRF-UNet | | 0.9823 | 0.4252 | 0.0356 | 0.1858 |
| **DMRF-UNet** | #1 | 0.9981 | 0.0674 | 0.1212 | 1.3869 |
| | #2 | **0.9845** | **0.3867** | **0.0317** | **0.1642** |

The comparison among the predicted permittivity distributions with one object buried in the heterogeneous soil is shown in Fig. 5. Figs. 5(a) and (b) show four examples of the input noisy B-scans and ground truths of permittivity distributions of the subsurface object, respectively. Figs. 5(c), (d), (e), and (g) show the predicted permittivity distributions from the PINet, Enc-Dec, U-Net, and SMRF-UNet, respectively. Fig. 5(h) presents the predicted results from the proposed DMRF-UNet, including the predicted denoised B-scan from MRF-UNet1 in Fig. 5(h.1), and the predicted permittivity distributions from MRF-UNet2 in Fig. 5(h.2). By comparing all these predicted results, the reconstructed permittivity distributions of the subsurface objects from the DMRF-UNet are more accurate than other works; especially the shapes of the objects are much closer to the ground truths.

Fig. 6 shows the inversion results when two separated objects are buried. Two-object scenarios are more challenging to be reconstructed than the one-object scenarios as the hyperbolic patterns from different objects may interfere with or disguise each other. As shown in Figs. 6(c)-(e), the networks in the existing studies cannot recover the permittivity distributions in these complex scenarios. Some of them even mistake the noise as the reflected signal from an extra object. As shown in Fig. 6(g), the SMRF-UNet recovers more accurate shapes and permittivity values of the objects than the previous networks, but still recognizes some noise as the target signal. Nevertheless, the proposed DMRF-UNet successfully extracts the main signatures [Fig. 6(h.1)] and accurately reconstructs the permittivity distributions of two separated objects [Fig. 6(h.2)]. Comparing with the results obtained using PINet, Enc-Dec, U-Net, and SMRF-UNet, the predicted permittivity distributions from the proposed scheme, shown in Fig. 6(h), are more precise regarding the shapes, sizes, positions, and permittivities of the subsurface objects.

Fig. 7 compares the inversion results when two subsurface objects are interfaced. In these cases, the hyperbolic patterns reflected from two objects are highly interleaved, making the reconstruction of the permittivity distribution even more difficult. As shown in Figs. 7(c)-(e), the PINet, Enc-Dec, and U-Net fail to differentiate two objects, and the predicted permittivity values greatly deviate from the ground truth. As shown in Fig. 7(g), the SMRF-UNet can roughly reconstruct the profiles of two interfaced objects, but the boundaries are quite blurred. On the contrary, the proposed network is capable of accurately reconstructing the permittivity distributions of the interfaced objects, as shown in Fig. 7(h). These results prove the effectiveness of the proposed scheme in accomplishing the inversion task with a higher resolution.

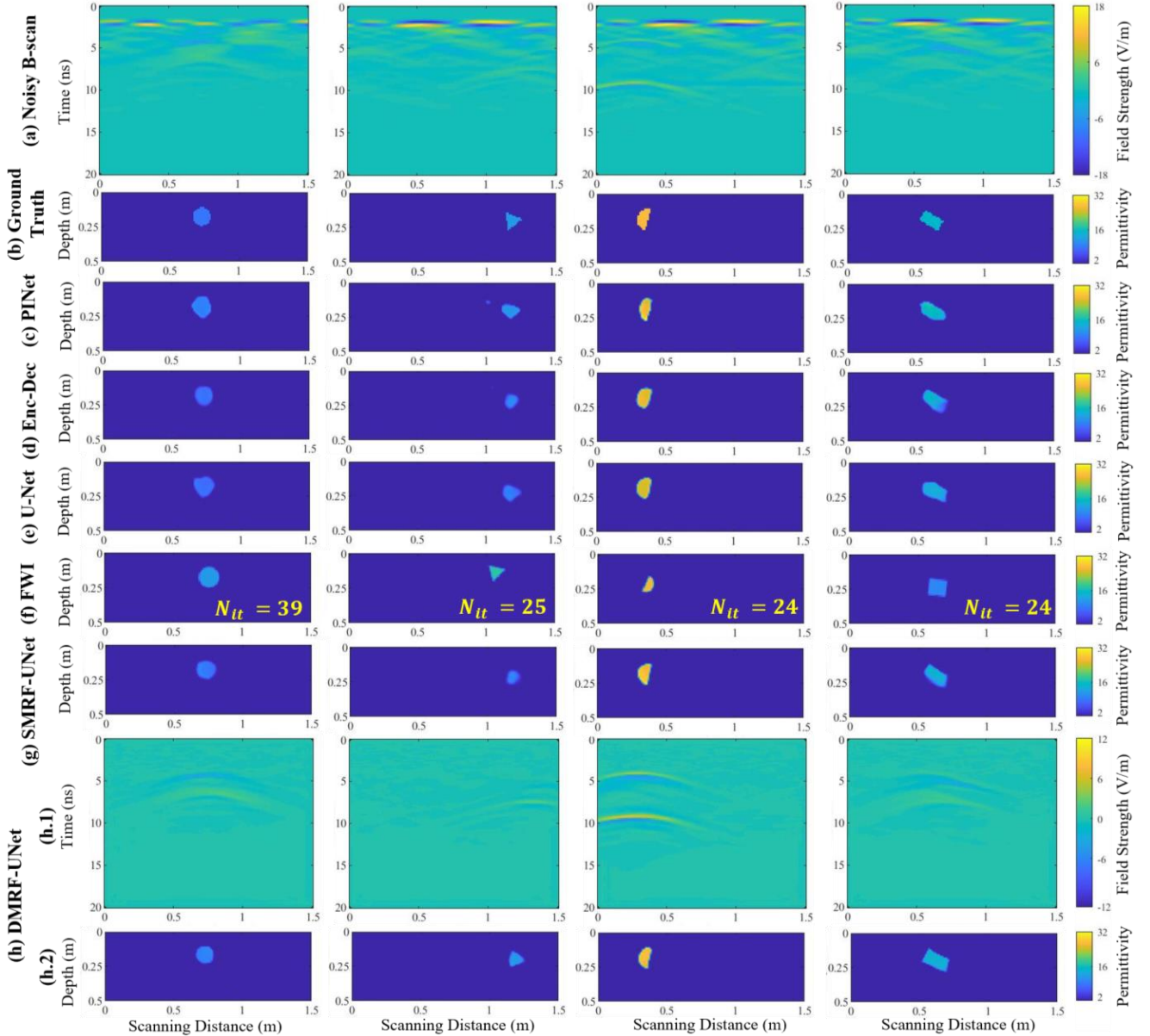

Fig. 5. Inversion results comparison when one object is buried. (a) Input noisy B-scans. (b) Ground truths of the permittivity distribution of subsurface objects. (c) Predicted permittivity distributions from the PINet. (d) Predicted permittivity distributions from the Enc-Dec. (e) Predicted permittivity distributions from the U-Net. (f) Reconstructed permittivity distributions using the FWI algorithm. The $N_{it}$ stands for the number of iterations required until the convergence of the optimization process. (g) Predicted permittivity distributions from the SMRF-UNet. (h) Predicted results from the proposed DMRF-UNet including (h.1) denoised B-scans from MRF-UNet1, and (h.2) permittivity distributions from MRF-UNet2. Note that the soil area (in the darkest blue) has a value of 0. The 'Permittivity' and 'Field Strength' represent the relative permittivity of the subsurface objects and the electric field strength, respectively.

### C. 2. Comparison with Conventional FWI Algorithm

To verify the superiority of the proposed scheme over traditional methods in reconstructing subsurface permittivity distributions from GPR B-scans, a comparative study with the widely used physics-driven FWI algorithm is performed. To apply the FWI algorithm to our cases, the 2D finite-difference time-domain gprMax simulator [27], [28] is used for forward modeling in each iteration [32], [33]. Simulated annealing optimization [34] is performed to minimize the MSE between the observed B-scan and the synthetic B-scan via updating the trial properties of subsurface objects. With the optimal object properties, the permittivity distribution of subsurface objects can be reconstructed for each observed B-scan. It should be noted that the aim of the FWI algorithm is to iteratively find the optimal permittivity distribution for a given B-scan, whereas deep learning-based methods are to find the optimal mapping relationship between B-scans and permittivity distributions.

Figs. 5-7(f) present the inversion results using the FWI algorithm for four one-object, two-separated-object, and two-interfaced-object scenarios, respectively. As shown in Fig. 5(f),

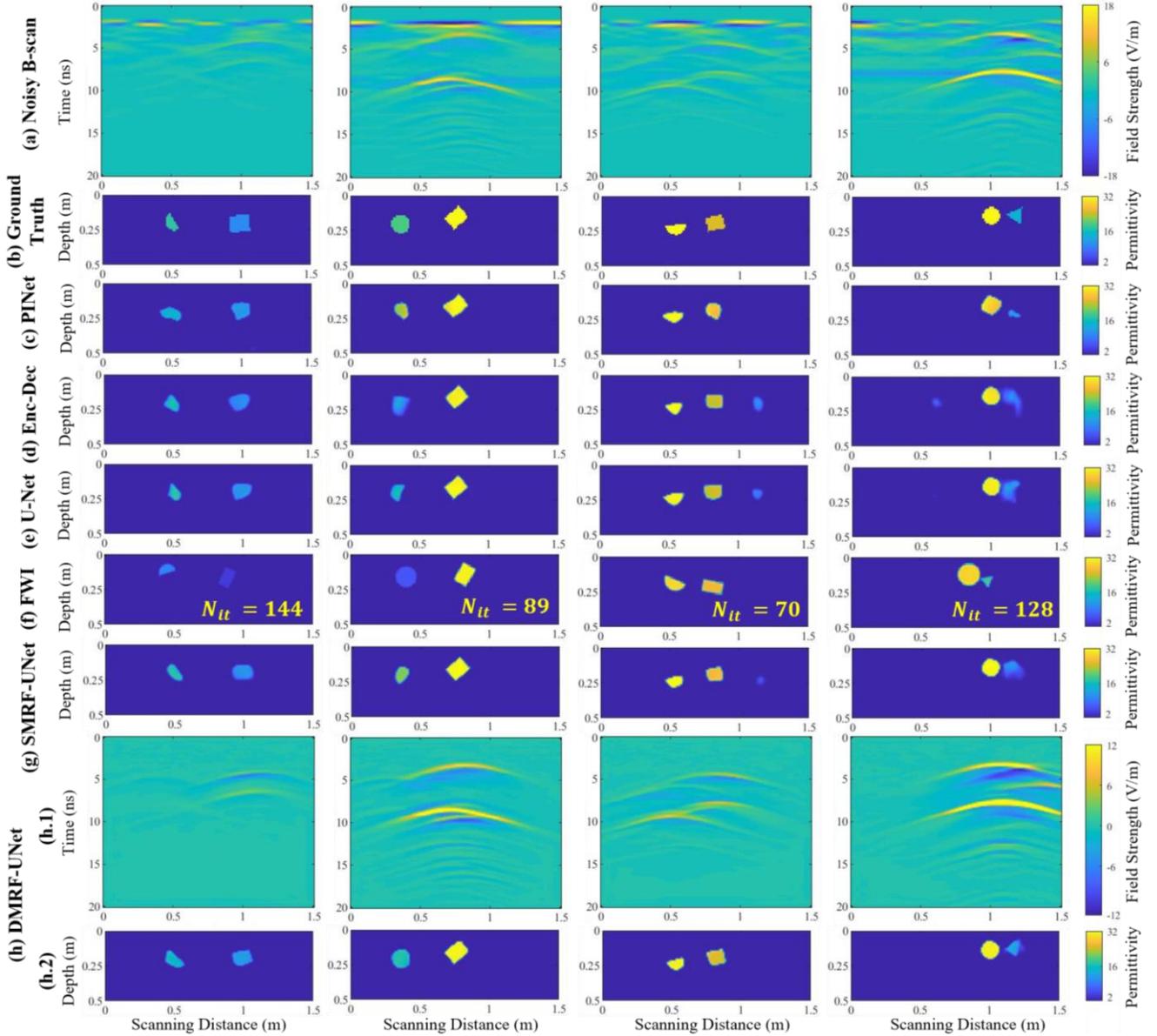

Fig. 6. Inversion results comparison when two separated objects are buried. (a) Input noisy B-scans. (b) Ground truths of the permittivity distribution of the subsurface objects. (c) Predicted permittivity distributions from the PINet. (d) Predicted permittivity distributions from the Enc-Dec. (e) Predicted permittivity distributions from the U-Net. (f) Reconstructed permittivity distributions using the FWI algorithm. (g) Predicted permittivity distributions from the SMRF-UNet. (h) Predicted results from the proposed DMRF-UNet.

FWI can roughly restore the permittivity distribution of one object, while the predicted sizes and shapes/orientations of the objects deviate from the ground truths. The number of iterations, $N_{it}$, required for convergence varies from 24 to 39. For two-separated-object scenarios shown in Fig. 6(f), the objects can be clearly observed in the images reconstructed using FWI, but some objects' permittivities and orientations are different from those in the ground truth images. For the two-interfaced-object cases shown in Fig. 7(f), the reconstructed profiles are far from their corresponding ground truths, especially regarding the objects' positions and orientations. As the updated variables in the optimization process for two-object scenarios are more than those for one-object scenarios, the required iteration number increases significantly, ranging from 70 to 151. Instead, the permittivity distributions reconstructed by the SMRF-UNet and DMRF-UNet are much closer to the ground truths, especially the DMRF-UNet that includes the denoising sub-network performs the best regarding the shapes, orientations, sizes, positions, and permittivities of the objects.

To quantitatively compare the performance of the FWI, SMRF-UNet, and DMRF-UNet, the evaluation metrics of the three types of scenarios shown in Figs. 5-7 are listed in Table II. As shown, all the FWI, SMRF-UNet, and DMRF-UNet perform better in one-object scenarios than in two-separated-object and two-interfaced-object scenarios, as the permittivity reconstruction of two objects is more challenging than that of one object. In each scenario type, the MSE, MAE and MRE values of DMRF-UNet are the lowest, and the SSIM is the highest, and those of SMRF-UNet are the second best, while the FWI performs the worst.

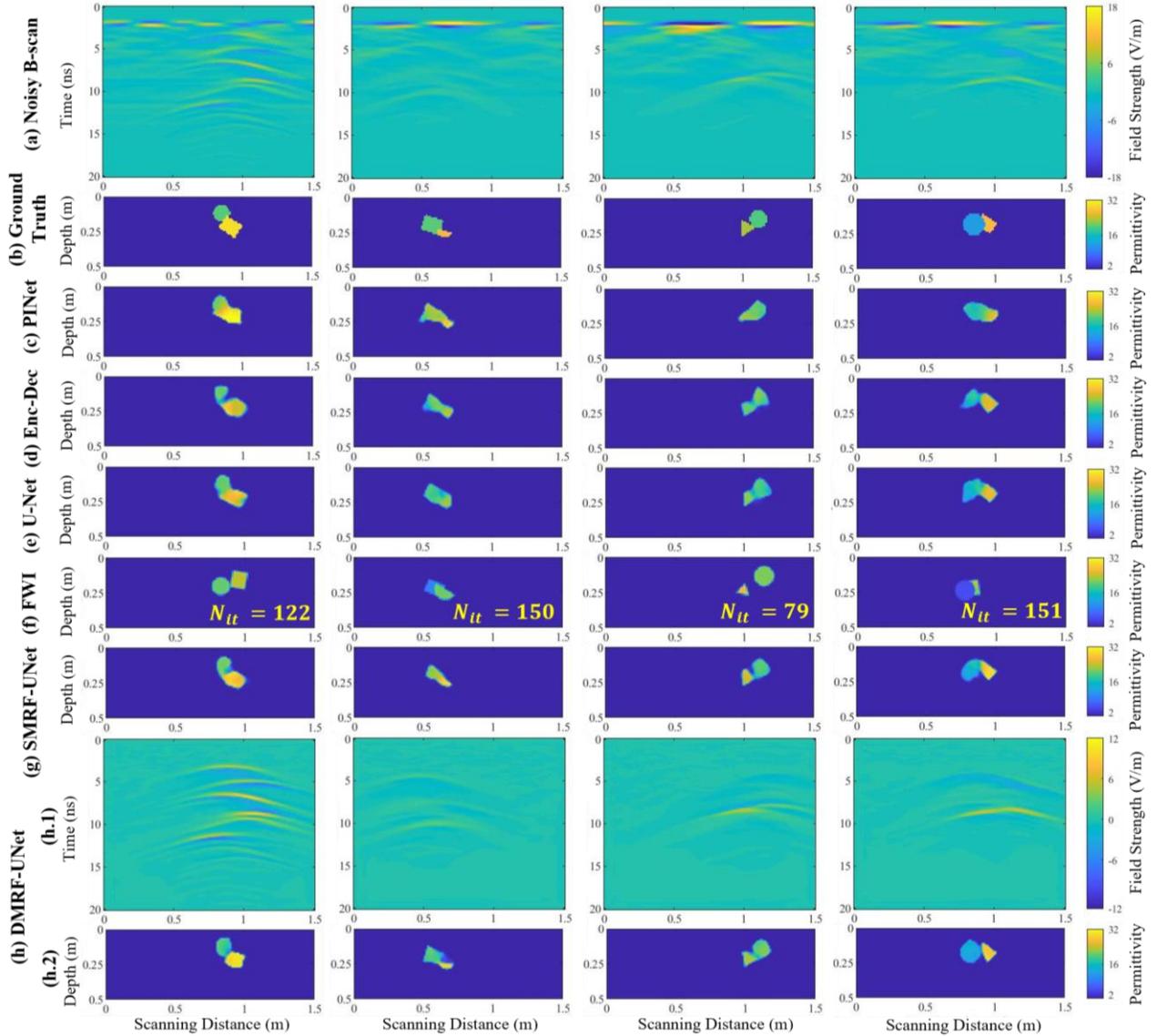

Fig. 7. Inversion results comparison when two interfaced objects are buried. (a) Input noisy B-scans. (b) Ground truths of the permittivity distribution of the subsurface objects. (c) Predicted permittivity distributions from the PINet. (d) Predicted permittivity distributions from the Enc-Dec. (e) Predicted permittivity distributions from the U-Net. (f) Reconstructed permittivity distributions using the FWI algorithm. (g) Predicted permittivity distributions from the SMRF-UNet. (h) Predicted results from the proposed DMRF-UNet.

TABLE II
COMPARISON WITH FWI ALGORITHM ON EVALUATION METRICS OF THE 12 TESTING DATA SHOWN IN FIGS. 5-7

| Scenario Type | MSE (↓) | | | MAE (↓) | | | MRE (%) (↓) | | | SSIM (↑) | | |
|---|---|---|---|---|---|---|---|---|---|---|---|---|
| | FWI | SMRF-UNet | DMRF-UNet | FWI | SMRF-UNet | DMRF-UNet | FWI | SMRF-UNet | DMRF-UNet | FWI | SMRF-UNet | DMRF-UNet |
| One Object | 1.5693 | 0.2225 | **0.1804** | 0.0950 | 0.0239 | **0.0196** | 0.6598 | 0.1736 | **0.1558** | 0.9709 | 0.9868 | **0.9875** |
| Two Separated Objects | 7.0984 | 0.8116 | **0.5544** | 0.3044 | 0.0675 | **0.0596** | 1.0541 | 0.2139 | **0.1847** | 0.9379 | 0.9693 | **0.9713** |
| Two Interfaced Objects | 3.8661 | 0.9687 | **0.8579** | 0.2089 | 0.0714 | **0.0571** | 0.7721 | 0.2656 | **0.2116** | 0.9590 | 0.9799 | **0.9816** |
| Ave. | 4.7119 | 0.6676 | **0.5298** | 0.2028 | 0.0543 | **0.0412** | 0.8287 | 0.2177 | **0.1840** | 0.9559 | 0.9787 | **0.9827** |

In terms of the computation time, the minimum number of iterations required by FWI in the considered scenarios is 24, and it takes approximately 1.5 hours to complete the inversion for one image. This is because the forward modeling is conducted in every iteration of the optimization process, and that takes about 3.5 minutes. For complicated scenarios such as the two-object cases, more iterations are required, resulting in higher computational cost. However, the well-trained DMRF-UNet is capable of reconstructing one image within 0.01 seconds. Although the network training is a computationally expensive process that takes about 16 hours, it only needs to be done once. Using the well-trained network, the permittivity distribution can be reconstructed from the input B-scan in near real-time.

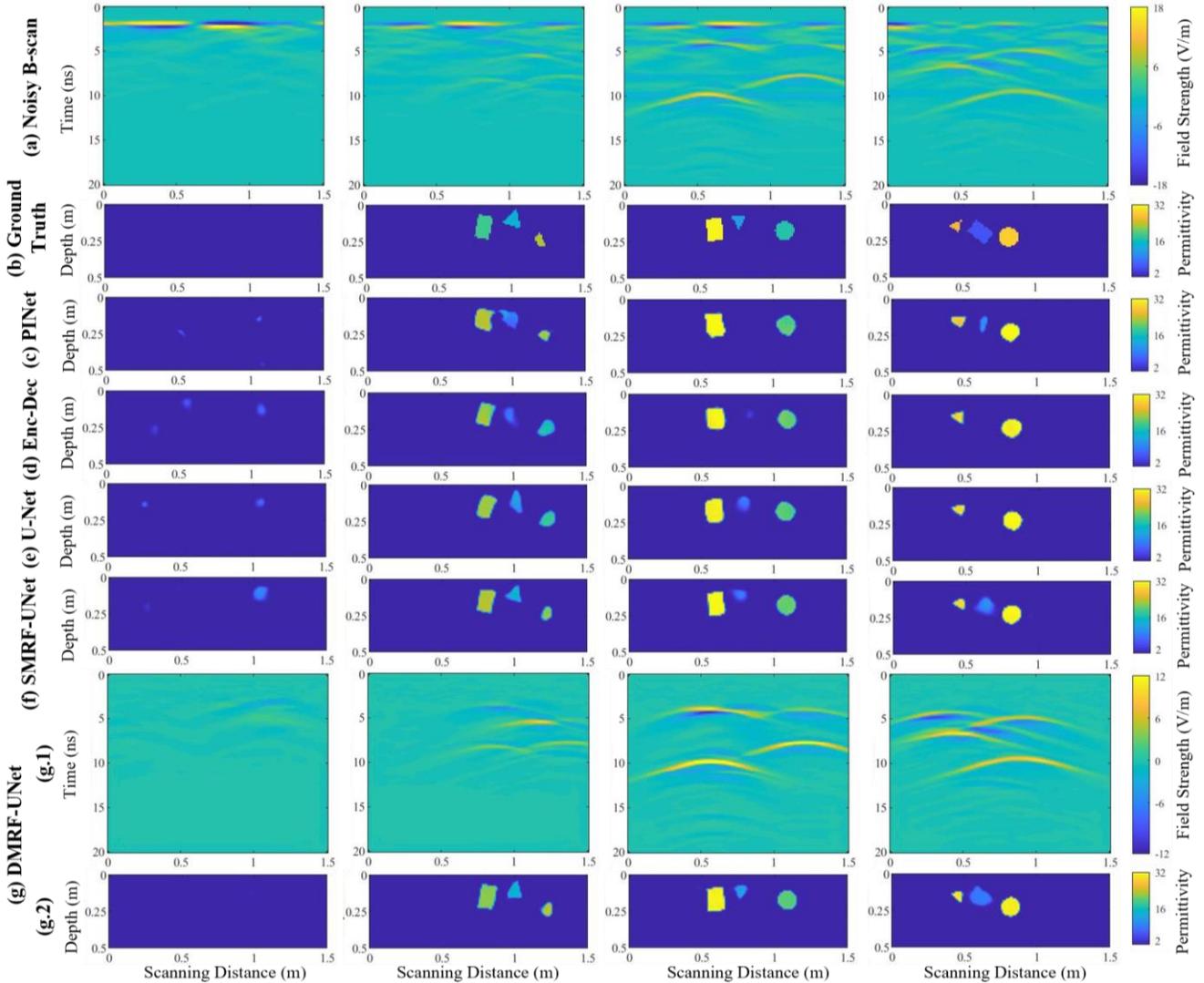

Fig. 8. Generalizability testing results when no object or three objects are buried. (a) Input noisy B-scans. (b) Ground truths of the permittivity distribution of the subsurface objects. (c) Predicted permittivity distributions from the PINet. (d) Predicted permittivity distributions from the Enc-Dec. (e) Predicted permittivity distributions from the U-Net. (f) Predicted permittivity distributions from the SMRF-UNet. (g) Predicted results from the proposed DMRF-UNet.

### D. Generalizability and Robustness Tests

#### D. 1. Zero/Three-Object Scenarios

TABLE III
COMPARISON ON EVALUATION METRICS OF ZERO/THREE-OBJECT SCENARIOS

| Network | | SSIM (↑) | MSE (↓) | MAE (↓) | MRE (%) (↓) |
|---|---|---|---|---|---|
| PINet [16] | | 0.9420 | 1.8482 | 0.1216 | 0.5236 |
| Enc-Dec [18] | | 0.9609 | 1.2172 | 0.0917 | 0.4138 |
| U-Net [18] | | 0.9623 | 1.1701 | 0.0875 | 0.3923 |
| SMRF-UNet | | 0.9649 | 1.0851 | 0.0878 | 0.3548 |
| **DMRF-UNet** | #1 | 0.9967 | 0.1429 | 0.2380 | 1.7041 |
| | #2 | **0.9691** | **0.9715** | **0.0764** | **0.3140** |

To evaluate the generalization capability of the proposed DMRF-UNet, 100 three-object and 10 new zero-object scenarios and their corresponding B-scans are produced and directly applied to the well-trained network without additional training. Note that only one-object and two-object scenarios are used in the training phase, and the distributions of the 10 zero-object scenarios are unseen for the network. The comparison with the existing networks and the SMRF-UNet on the evaluation metrics of these 110 scenarios is provided in Table III. For the MREs, only the 100 three-object scenarios are considered as the $max(|y_{i,j}|)$ (in Eqn. (7)) in the 10 zero-object scenarios are always 0. As shown, the SSIM, MSE, MAE, and MRE of the inversion results using the proposed DMRF-UNet are much better than those of PINet, Enc-Dec, U-Net, and SMRF-UNet.

The imaging result comparison for the generalized scenarios is presented in Fig. 8. For the zero-object scenarios, the reconstructed maps using the proposed DMRF-UNet perfectly match with the ground truths. However, Enc-Dec, PINet, U-Net, and SMRF-UNet mistakenly identify the noise and clutters as the objects' signatures and reconstruct some fuzzy objects. When three objects are considered, PINet, Enc-Dec, U-Net, and SMRF-UNet can only restore two of the objects or reconstruct a fuzzy map of three objects, especially for the objects with closer permittivities to the soil. In contrast, the DMRF-UNet is still capable of denoising the B-scan [Fig. 8(g.1)] and rebuilds the permittivity distribution of the three objects with various properties [Fig. 8(g.2)]. These comparative results demonstrate the better generalization capability of the DMRF-UNet in diverse subsurface scenarios over the previous networks.

TABLE IV
PROPERTIES OF THE UTILIZED HETEROGENEOUS SOIL MODELS AND EVALUATION METRICS COMPARISON

| Soil Model | Sand Fraction | Clay Fraction | Water Content | Number of Materials | Permittivity Range | Conductivity Range | SSIM (↑) #1 | #2 | MSE (↓) #1 | #2 | MAE (↓) #1 | #2 | MRE (%) (↓) #1 | #2 |
|---|---|---|---|---|---|---|---|---|---|---|---|---|---|---|
| (1) | 0.7 | 0.3 | 0.1%-30% | 10 | 4.60-13.15 | 0.03-0.07 | 0.9961 | 0.9758 | 0.1881 | 0.6274 | 0.3370 | 0.0552 | 2.9823 | 0.2769 |
| (2) | 0.6 | 0.4 | 0.1%-25% | 15 | 4.08-11.57 | 0.02-0.07 | 0.9968 | 0.9816 | 0.1356 | 0.4152 | 0.2449 | 0.0373 | 1.3733 | 0.2440 |
| Pre-Trained | 0.5 | 0.5 | 0.1%-20% | 20 | 3.82-9.99 | 0.01-0.07 | 0.9981 | 0.9845 | 0.0674 | 0.3867 | 0.1212 | 0.0317 | 1.3869 | 0.1642 |
| (3) | 0.4 | 0.6 | 0.1%-15% | 25 | 3.64-8.36 | 0.01-0.06 | 0.9968 | 0.9840 | 0.1289 | 0.3247 | 0.2329 | 0.0328 | 1.4743 | 0.1643 |
| (4) | 0.3 | 0.7 | 0.1%-10% | 30 | 3.57-6.72 | 0.01-0.05 | 0.9966 | 0.9822 | 0.1706 | 0.2996 | 0.3003 | 0.0311 | 1.6429 | 0.1870 |

*D. 2. New Soil Environments*

To evaluate the robustness of the proposed method in reconstructing permittivity distributions under new heterogeneous soil conditions, four new Peplinski mixing models are used to test the performance of the DMRF-UNet. Table IV lists the parameters of these soil models, where the network with the soil model described in Section III.A is listed as "pre-trained", and the soil models (1)-(4) represent four soil environments that are totally new. As supervised deep learning-based techniques are data-driven, it is challenging to directly apply the well-trained network to a new dataset that is vastly different from the training dataset. To adapt the pre-trained network to new datasets with the soil models (1)-(4), transfer learning is introduced [35]. In particular, a small dataset with the new soil model is produced to fine-tune the pre-trained network. Every dataset includes 360 sets of noisy B-scans, denoised B-scans, and permittivity distributions of subsurface objects. In the fine-tuning process, a small initial learning rate of 0.0001 is used to avoid overfitting. The learning rate is reduced to 99% of its value if the training loss has no drop in the last epoch. After 50-epoch training, the fine-tuned network is used to predict the permittivity distributions from 36 testing B-scans obtained for the new soil models.

As shown in Table IV, the SSIMs, MSEs, MAEs, and MREs of the denoising (#1) and inversion (#2) for the four new soil models and the pre-trained model are compared. For the denoising, the performance of the soil models (2)-(3) is better than the models (1) and (4), as the properties of the soil models (2)-(3) are closer to the one that is well pre-trained using a large diverse set of data. Among these soil models, the soil model (1) performs the worst but yet with high SSIM and low MSE, MAE, and MRE. For the inversion, the SSIMs of the soil models (1)-(2) are lower and the MSEs, MAEs, MREs are higher than those of the pre-trained model, but the results are still satisfactory. The performance degradation is due to the increase of the soil water content ratio, resulting in a rise of the reflected signal attenuation, which makes the reconstruction of the subsurface permittivity distribution more challenging. For the soil models (3)-(4) with lower signal attenuations, the metrics are quite close to the results of the pre-trained model, the model (4) even slightly outperforms the pre-trained one.

Fig. 9 presents the inversion results of the fine-tuned network with the four heterogeneous soil models. Although the clutters and noise in the noisy B-scans [Fig. 9(a)] are vastly different from those in the pre-trained set, the hyperbolic signatures of objects are still clearly extracted by MRF-UNet1, as shown in Fig. 9(c). After that, the permittivity distributions are reconstructed by MRF-UNet2. As shown in Fig.9(d), the predicted maps with the soil models (1)-(4) match well with their ground truths [Fig. 9(b)]. These testing results demonstrate that even when the soil environment is totally new for the network, the permittivity distributions can still be reconstructed with high accuracy via fine-tuning the pre-trained network using an additional small set of new data.

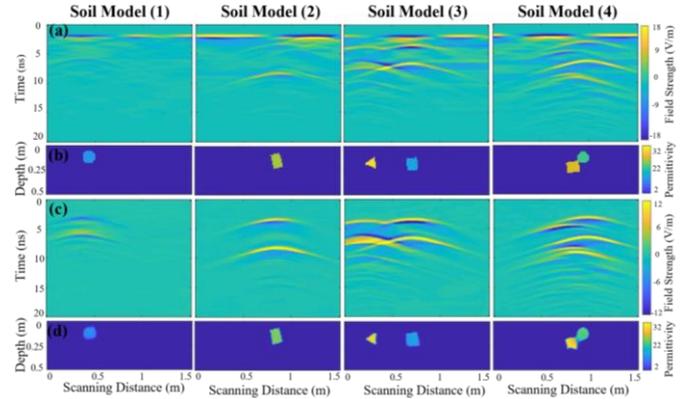

Fig. 9. Inversion results with four new soil models after fine-tuning the pre-trained model using small sets of new data. (a) Input noisy B-scans. (b) Ground truths. (c) Predicted denoised B-scans. (d) Predicted permittivity distributions.

*D. 3. New Objects*

To further test the generalization capability of the proposed method for new objects via transfer learning, two small datasets for the objects with (1) new permittivities and (2) new profiles are generated. In the dataset (1), the object's permittivity is randomly selected from [32, 42]. In the dataset (2), to model random shapes of subsurface objects that are very different from those in the pre-training set, the handwritten '1'-shaped objects in the MNIST database [36] are randomly rotated and buried in the soil, as shown in Fig. 10. Every dataset includes 360 sets of noisy B-scans, denoised B-scans, and permittivity distributions for fine-tuning the pre-trained network described in Section III.A. Other parameter settings are the same as in Section III.D.2. The fine-tuned network predicts the permittivity distributions of the objects with new permittivities or new profiles from 36 testing B-scans.

Table V lists the evaluation metrics of the two sets of new testing data. Compared to the results of the pre-trained model, the denoising and inversion performance for the objects with new permittivities and new profiles is slightly degraded but satisfactory. Fig. 10 shows the imaging results for one/two objects with new permittivities and new profiles. The predicted permittivity distributions shown in Fig. 10(d) match well with their corresponding ground truths shown in Fig. 10(b). These results demonstrate that when the properties of subsurface objects are out of the pre-training set, the permittivity distributions of the new objects are still accurately reconstructed via the fine-tuned network.

TABLE V
METRICS COMPARISON FOR OBJECTS WITH PERMITTIVITIES AND PROFILES

| Models | | SSIM (↑) | MSE (↓) | MAE (↓) | MRE (%) (↓) |
|---|---|---|---|---|---|
| Pre-Trained | #1 | 0.9981 | 0.0674 | 0.1212 | 1.3869 |
| | #2 | 0.9845 | 0.3867 | 0.0317 | 0.1642 |
| (1) New Permittivities | #1 | 0.9918 | 0.1211 | 0.1911 | 1.0865 |
| | #2 | 0.9879 | 1.7613 | 0.0766 | 0.2010 |
| (2) New Profiles | #1 | 0.9883 | 0.0672 | 0.1432 | 1.8047 |
| | #2 | 0.9827 | 0.9514 | 0.0712 | 0.3319 |

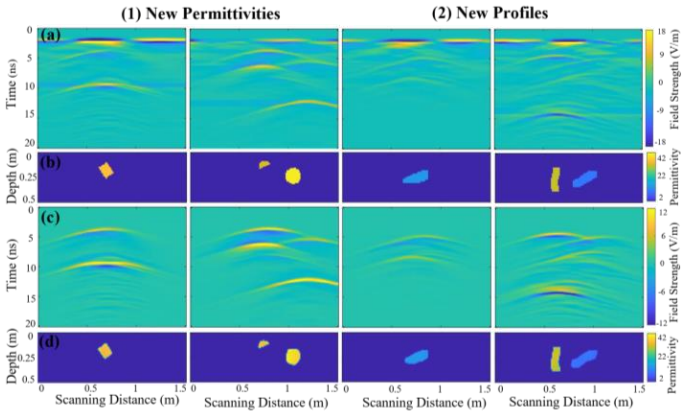

Fig. 10. Inversion results for objects with (1) new permittivities and (2) new profiles. (a) Input noisy B-scans. (b) Ground truths. (c) Predicted denoised B-scans. (d) Predicted permittivity distributions.

*E. Ablation Study*

To prove the effectiveness of the denoising sub-network MRF-UNet1, the MRF module, the end-to-end training method, and the hybrid two-channel input of MRF-UNet2, an ablation study is conducted. We compare the performance of the networks when (i) the MRF-UNet1 for signature extraction is removed, which is equivalent to the SMRF-UNet, (ii) the MRF module is replaced by regular convolutions, (iii) the end-to-end learning is replaced by the separated training, which trains the MRF-UNet1 firstly and then trains the MRF-UNet2, (iv) the hybrid input of MRF-UNet2 is replaced by the one-channel denoised B-scan, and (v) the final model combines all the proposed components. The comparative metrics are presented in Table VI. As shown, the absence of MRF-UNet1 in (i) leads to the lower SSIM and the higher MSE, MAE, and MRE. This performance degradation is because the noise and clutters in the B-scan under heterogeneous soil conditions interfere the identification of object signatures. The lack of MRF module in (ii) reduces the feature extraction capability in the convolutions, which results in a decrease in SSIM and an increase in MSE, MAE, and MRE. The separate training strategy in (iii) reduces the inversion accuracy compared to the end-to-end training method, which balances the training accuracy of the two sub-networks and compensates for the information loss in predicting the denoised B-scan from MRF-UNet1. The one-channel denoised input in (iv) results in lower inversion accuracy than the hybrid two-channel input, as information loss occurs in the denoising process of MRF-UNet1. The addition of the noisy B-scan in the hybrid two-channel input effectively avoids the information loss in the first-stage denoising process and improves the second-stage inversion performance. Overall, implementing all the components in (v) (the proposed network architecture) leads to the best results with the highest SSIM and the lowest MSE, MAE, and MRE.

TABLE VI
ABLATION STUDY

| Study/Metrics | (i) | (ii) | (iii) | (iv) | (v) |
|---|---|---|---|---|---|
| Signature Extraction | ✗ | ✓ | ✓ | ✓ | ✓ |
| MRF Module | ✓ | ✗ | ✓ | ✓ | ✓ |
| End-to-End Training | ✓ | ✓ | ✗ | ✓ | ✓ |
| Two-Channel Input | N.A. | ✓ | ✓ | ✗ | ✓ |
| SSIM (↑) | 0.9823 | 0.9790 | 0.9818 | 0.9844 | **0.9845** |
| MSE (↓) | 0.4252 | 0.4365 | 0.4620 | 0.4008 | **0.3867** |
| MAE (↓) | 0.0356 | 0.0380 | 0.0381 | 0.0327 | **0.0317** |
| MRE (%) (↓) | 0.1858 | 0.1946 | 0.1938 | 0.1685 | **0.1642** |

IV. TESTS WITH REAL MEASUREMENT DATA

*A. Measurement Data Collection and Network Training*

The performance of the proposed DMRF-UNet in reconstructing subsurface permittivity distributions is further examined using the field-measured GPR data. As shown in Figs. 11(a)-(c), the measurement with a commercial GSSI's Utility Scan Pro GPR system is conducted on an outdoor uneven sandy field. Fig. 11(a) presents the schematic view of the experiment scenario. As illustrated in Fig. 11(b), the GPR system with a 400-MHz antenna and a control unit is used to collect B-scans in the experiments. Every B-scan obtained along a one-meter scanning trace consists of 88 A-scans, and the number of sampling points in every A-scan is 512. The time window is 20 ns. The buried objects are selected as five wooden objects with various shapes, sizes, and relative permittivities ($\varepsilon_r$). The properties of the buried objects are shown in Fig. 11(c). The ranges of the $x$-coordinate position, depth, horizontal angle, and vertical angle of the object are selected from [20, 80] cm, [9, 25] cm, [0, 60] degree, and [0, 60] degree, respectively. The permittivity distribution of the field surface is measured by Keysight N1501A Dielectric Probe. As shown in Fig. 11(d), the relative permittivity along the scanning trace varies from 2.64 to 6.48 due to the difference in humidity.

The measured B-scans are normalized to [0, 1], resized to 128×128, and pre-processed by mean subtraction [22] to form the input noisy B-scans. To obtain the ground truth of the denoised B-scan, a B-scan is measured in an empty area far away from the object region and subtracted from the noisy B-scan. Totally 180 sets of data, including the measured noisy B-scans, the denoised B-scans, and the corresponding permittivity distributions of subsurface objects, are obtained to fine-tune the network pre-trained by simulation data as described in Sections III.A-B. The initial learning rate is set as 0.0001, and then decreased to 99% of its value to avoid over-fitting if the training loss does not decrease in the last epoch.

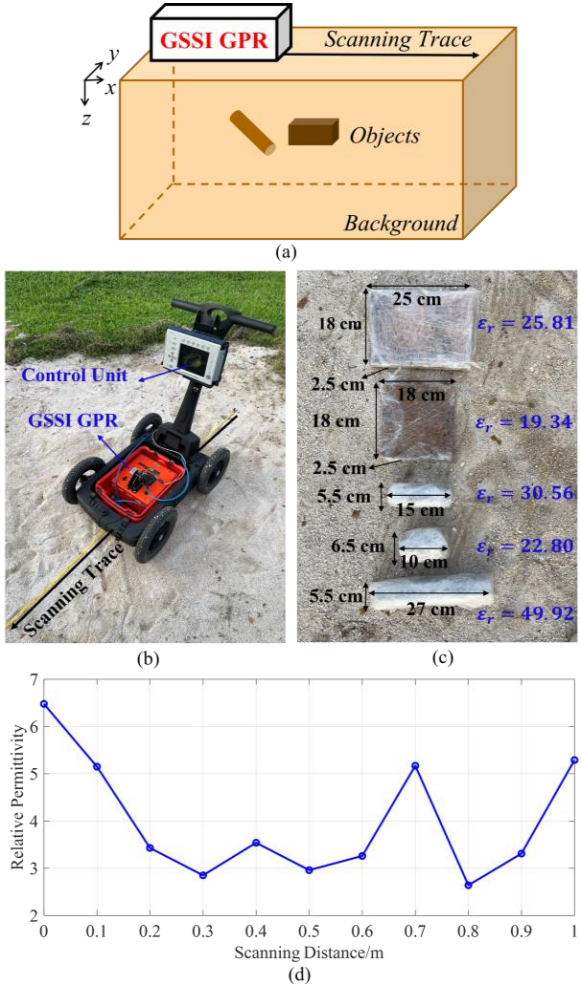

Fig. 11. The experiment setup for collecting real measurement data. (a) The schematic view of the experiment scenario. (b) The view of the real scenario with the commercial GPR system. (c) Five wooden objects utilized in the experiments. (d) The permittivity distribution of the field surface.

### B. Inversion Result Comparison and Analysis

After 200-epoch training, four field-measured noisy B-scans are used to test the fine-tuned network. To quantitatively evaluate the performance of the proposed DMRF-UNet, the comparison with the FWI algorithm, the existing inversion networks and the SMRF-UNet, which are fine-tuned with transfer learning as well, on evaluation metrics is shown in Table VII. It can be observed that the accuracy of the denoising (#1) and inversion (#2) using the DMRF-UNet is still high. The SSIM, MSE, MAE, and MRE of the denoising reach 0.9820, 1.3265, 0.7082, 1.0962%. For the task of inversion, the MSE (0.3881), MAE (0.0537), and MRE (0.1914%) of DMRF-UNet are much lower than all the results of FWI, PINet, Enc-Dec, U-Net, and SMRF-UNet, while the SSIM (0.9589) is slightly lower than that of PINet. Overall, the performance of the proposed DMRF-UNet on the real measurement data is the best among those of the previous studies.

To qualitatively compare the inversion accuracy, the imaging results of the PINet, Enc-Dec, U-Net, FWI, SMRF-UNet, and DMRF-UNet are compared in Fig. 12. Figs. 12(a)-(b) present the measured noisy B-scans and the corresponding ground truths of the permittivity distributions of subsurface objects. As shown in Figs. 12(c)-(g), in Case 1, the PINet, Enc-Dec, U-Net, FWI, and SMRF-UNet roughly restore the size, shape, and position of one circular object but with an inaccurate permittivity value. In Case 2, the PINet, Enc-Dec, U-Net, and SMRF-UNet fail to reconstruct the rectangle object's permittivity distribution, and the FWI cannot accurately restore the object's permittivity and location. In Cases 3-4, the two-separated-object and two-interfaced-object scenarios, the inversion results of PINet, Enc-Dec, U-Net, and FWI deviate dramatically from the ground truths, and the SMRF-UNet cannot precisely restore the shapes and permittivity values especially for the interfaced objects. In contrast, as shown in Fig. 12(h.2), the reconstructed permittivity distributions of the proposed DMRF-UNet in Cases 1-4 agree well with the ground truths. Although some objects' boundaries are slightly blurred, the sizes, shapes, positions, and permittivities of the subsurface objects are restored with high accuracy. This is because the DMRF-UNet effectively extracts the object reflections from the noisy B-scans, as shown in Fig. 12(h.1), to alleviate the adverse effect of environmental factors on the reconstruction accuracy, while the previous networks without denoising cannot distinguish the objects' reflection signatures and the environmental noise well. These quantitative and qualitative results demonstrate the superior performance of the DMRF-UNet compared to the previous methods on the field-measured data. The good agreement between the predicted and the actual permittivity distributions indicates the effectiveness of the proposed methodology for real field applications.

TABLE VII
COMPARISON ON EVALUATION METRICS OF REAL MEASUREMENT DATA

| Method | | SSIM (↑) | MSE (↓) | MAE (↓) | MRE (%) (↓) |
|---|---|---|---|---|---|
| FWI | | 0.9412 | 10.5406 | 0.3449 | 0.8371 |
| PINet | | **0.9654** | 2.4186 | 0.1503 | 0.5290 |
| Enc-Dec | | 0.9478 | 0.9227 | 0.0932 | 0.3456 |
| U-Net | | 0.9506 | 0.6564 | 0.0820 | 0.3135 |
| SMRF-UNet | | 0.9579 | 0.5044 | 0.0786 | 0.3037 |
| **DMRF-UNet** | #1 | 0.9820 | 1.3265 | 0.7082 | 1.0962 |
| | #2 | 0.9589 | **0.3881** | **0.0537** | **0.1914** |

## V. CONCLUSION

This work proposes a two-stage deep learning-based method for GPR data inversion under heterogeneous soil conditions. The first stage employs a U-shape DNN to extract the object signatures in the noisy B-scan and remove the noise and clutters. The second stage takes the concatenated denoised image and noisy image as input, and reconstructs the permittivity distributions of subsurface objects. Convolutions with multiple receptive fields are introduced in both stages to extract multi-scale features from the B-scans. An end-to-end training method is proposed by combining the losses of two stages. The test results demonstrate that the proposed network allows a robust and accurate reconstruction of the permittivity distributions of subsurface objects from the GPR B-scans. The comparison with existing techniques shows the superiority of the proposed scheme, especially for the complex scenarios where multiple objects are involved. This study aims to address the challenges of GPR inverse problems under heterogeneous soil conditions using deep learning techniques. Our future work will focus on the investigation of novel network architectures for the 3D permittivity distribution reconstruction.

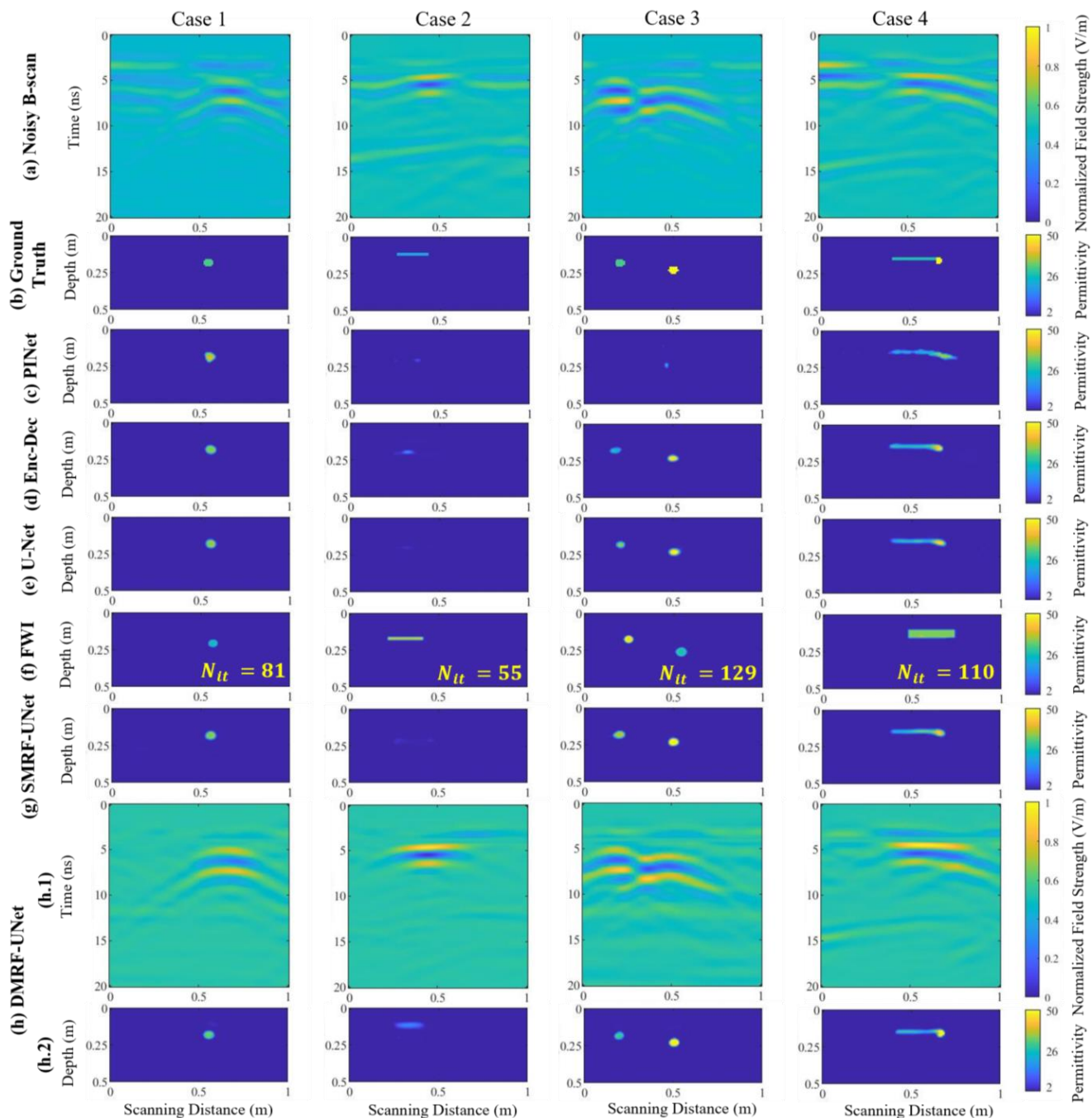

Fig. 12. Inversion result comparison of real measurement data. (a) Input noisy B-scans. (b) Ground truths of the permittivity distribution of the subsurface objects. (c) Predicted permittivity distributions from the PINet. (d) Predicted permittivity distributions from the Enc-Dec. (e) Predicted permittivity distributions from the U-Net. (f) Reconstructed permittivity distributions using the FWI algorithm. (f) Predicted permittivity distributions from the SMRF-UNet. (h) Predicted results from the proposed DMRF-UNet including (h.1) denoised B-scans from MRF-UNet1, and (h.2) permittivity distributions from MRF-UNet2.

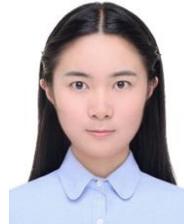
**Qiqi Dai** received the B.Eng. degree in communication engineering from Soochow University, Suzhou, China, in 2018, and the M.Sc. degree from the School of Electrical and Computer Engineering, National University of Singapore, Singapore, in 2019. She is currently pursuing the Ph.D. degree in the School of Electrical and Electronic Engineering, Nanyang Technological University, Singapore. Her research interests include the ground penetrating radar and deep learning-based image processing techniques.

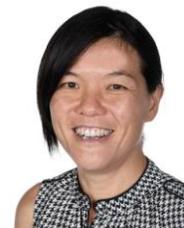
**Yee Hui Lee** (Senior Member, IEEE) received the B.Eng. (Hons.) and M.Eng. degrees from the School of Electrical and Electronics Engineering, Nanyang Technological University, Singapore, in 1996 and 1998, respectively, and the Ph.D. degree from the University of York, York, U.K., in 2002. Since 2002, she has been a Faculty Member with Nanyang Technological University, where she is currently an Associate Professor with the School of Electrical and Electronic Engineering. Her research interests include the channel characterization, rain propagation, antenna design, electromagnetic bandgap structures, and evolutionary techniques.

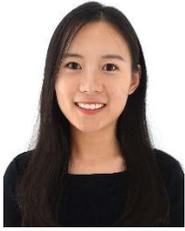

**Hai-Han Sun** received the bachelor's degree in electronic information engineering from the Beijing University of Posts and Telecommunications (BUPT), Beijing, China, in 2015, and the Ph. D. degree in engineering from the University of Technology Sydney (UTS), Australia, in 2019. She is currently a Research Fellow with the School of Mechanical and Aerospace Engineering, Nanyang Technological University, Singapore. Her research interests include base station antennas and ground penetrating radar.

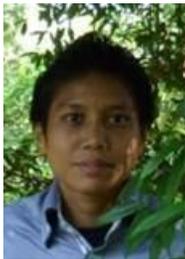

**Genevieve Ow** received the B.S. degree (Hons.) in plant biology from Massey University, New Zealand, and the Ph.D. degree in plant biology from University of Canterbury, New Zealand. She is currently a deputy director at the Centre for Urban Greenery & Ecology, National Parks Board, Singapore. Her research interests include ecophysiology of resource acquisition in urban ecosystems, tree biology/arboriculture, urban sustainability, plant physiology, and responses of plants to extremes of environment.

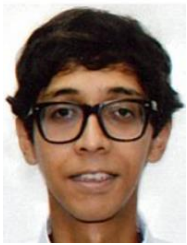

**Mohamed Lokman Mohd Yusof** received the B.S. degree in biomedical science from the University of Western Australia. He is presently an executive at the Centre for Urban Greenery and Ecology, National Parks Board and manage research projects in the field of arboriculture, remote sensing, and phytoremediation.

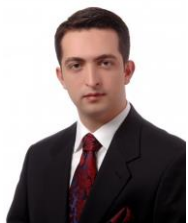

**Abdulkadir C. Yucel** (Senior Member, IEEE) received the B.S. degree in electronics engineering (Summa Cum Laude) from Gebze Institute of Technology, Kocaeli, Turkey, in 2005, and the M.S. and Ph.D. degrees in electrical engineering from the University of Michigan, Ann Arbor, MI, USA, in 2008 and 2013, respectively.

From September 2005 to August 2006, he worked as a Research and Teaching Assistant at Gebze Institute of Technology. From August 2006 to April 2013, he was a Graduate Student Research Assistant at the University of Michigan. Between May 2013 and December 2017, he worked as a Postdoctoral Research Fellow at various institutes, including the Massachusetts Institute of Technology. Since 2018, he has been working as an Assistant Professor at the School of Electrical and Electronic Engineering, Nanyang Technological University, Singapore.

Dr. Yucel received the Fulbright Fellowship in 2006, Electrical Engineering and Computer Science Departmental Fellowship of the University of Michigan in 2007, and Student Paper Competition Honorable Mention Award at IEEE International Symposium on Antennas and Propagation in 2009. He has been serving as an Associate Editor for the International Journal of Numerical Modelling: Electronic Networks, Devices and Fields and as a reviewer for various technical journals.